\documentclass[traditabstract]{aa} 
\usepackage{graphicx}
\usepackage{txfonts}
\usepackage{natbib}
\bibpunct{(}{)}{;}{a}{}{,}
\usepackage[breaklinks=true]{hyperref}
\hypersetup{colorlinks=true,citecolor=blue,linkcolor=blue,filecolor=blue,runcolor=blue}
\usepackage{caption}

\begin{document}
   \title{The accretion environment in Vela X-1 during a flaring period using XMM-Newton}

  \author{S. Mart\'inez-N\'u\~nez
          \inst{1}
          \and
	  J.M. Torrej\' on\inst{1,2}
	  \and
	  M. K\"{u}hnel\inst{3}
	  \and
	  P. Kretschmar\inst{4}
	  \and 
	  M. Stuhlinger\inst{4}
          \and
	  J.J.Rodes-Roca\inst{1,2}
	  \and
	  F. F\"{u}rst\inst{5}
	  \and\\
	  I. Kreykenbohm\inst{3}
	  \and
	  A. Martin-Carrillo\inst{6}
	  \and
	   A.M.T. Pollock\inst{4}
	  \and
	  J. Wilms\inst{3}
          }
 
    \institute{Instituto Universitario de F\'isica Aplicada a las Ciencias y las Tecnolog\'ias, \\University of Alicante, P.O. Box 99,
     E03080 Alicante, Spain \email{silvia.martinez@ua.es}
    \and
     Departamento de F\'isica, Ingener\'ia de Sistemas y Teor\'ia de la Se\~{n}al, University of Alicante, P.O. Box 99,
     E03080 Alicante, Spain
    \and
    Dr. Karl Remeis-Observatory \& ECAP, Universitat Erlangen-N\"urnberg, Sternwartstr. 7, D96049 Bamberg, Germany
    \and
    European Space Astronomy Centre (ESA/ESAC), Science Operations Department, Villanueva de la Ca\~{n}ada (Madrid), Spain
    \and
    Cahill Center for Astronomy and Astrophysics, California Institute of Technology, Pasadena, CA 91125, USA
    \and
    University College Dublin, Belfield, Dublin 4 Ireland}

   \date{Received July 30, 2013, accepted December 26, 2013}
 
  \abstract {We present analysis of 100~ks contiguous XMM-Newton data
of the prototypical wind accretor Vela X-1. The observation covered 
eclipse egress between orbital phases 0.134 and 0.265,  during which a 
giant flare took place, enabling us to study the spectral properties both
outside and during the flare. This giant flare with a peak luminosity 
of $3.92^{+0.42}_{-0.09} \times 10^{37}$~erg~s$^{-1}$ allows 
estimates of the physical parameters of the accreted structure 
with a mass of $\sim$ $10^{21}$~g.

We have been able to model several contributions 
to the observed spectrum with a phenomenological model
formed by three absorbed power laws plus three emission lines. 
After analysing the variations with orbital phase of the column density of 
each component, as well as those in the Fe and Ni fluorescence lines, we 
provide a physical interpretation for each spectral component. Meanwhile, the
first two components are two aspects of the principal accretion component
from the surface of the neutron star, the third component
seems to be the \textit{X-ray light echo} formed
in the stellar wind of the companion.

}
  
\maketitle
%

\section{Introduction}

Vela X-1 is an eclipsing high-mass X-ray binary discovered in 1967 \citep{chodil67} 
at a distance of $\sim$ 2~kpc \citep{nagase89}. It consists of the early type 
supergiant HD~77581 (B0.5Iab, R $\sim$ 34 R$_\odot$) and a neutron star of
$\sim$ 1.77M$_\odot$ \citep{rawls2011}, orbiting its companion 
with a period of 8.964 days. The distance at periastron is 
only 0.6 stellar radii (a $\sim$ 53 R$_\odot$) 
\citep{vanKer95} from the stellar surface, well within the 
acceleration zone of the stellar wind. Despite the status of the Vela X-1 
system as an archetypal wind accretor, the structure of the extended atmosphere 
of the supergiant and large scale structure of the stellar wind 
are not known in detail.

The X-ray luminosity of Vela X-1 is typically a few $10^{36}$~erg~s$^{-1}$,
four or more orders of magnitude brighter than the intrinsic emission expected from 
the supergiant primary. It is well explained by wind accretion from the optical companion, with an
estimated mass loss rate of 2$\times$ 10$^{-6}$ M$_\odot$  yr$^{-1}$ \citep{Watanabe06}.
Intense X-ray variability of Vela X-1, lasting from a few minutes to several hours,
is commonly thought to be associated with changes in the accretion rate and to inhomogeneities
in the stellar wind flow \citep{nagase83,haberl90}. There have been reports,
on the one hand, of reductions in the flux to less than 10$\%$ of its normal emission 
(``off states''), as well as of very active states, on the other,
with increases in the flux by sometimes more than a factor of 10 \citep{krey99,krey08}.

X-ray pulsations were discovered by \citet{McClin76} with a
fluctuating pulse period of $\sim$ 283 s. This random spin up 
and down does not show any apparent connection to the orbital phase 
of the neutron star around its supergiant companion \citep{nagase84},
and it is believed to be caused by the transfer of angular momentum by wind 
accretion onto the neutron star.

The X-ray spectrum has been described either by a power law with
a high energy cut-off beyond 15-30 keV \citep{nagase86,krey99}
or by a negative positive exponential model (NPEX) 
\citep{orlandini98,krey02,odaka13}. Below 3 keV, a soft excess is observed, 
which was modelled by thermal Bremsstrahlung with kT $\sim$ 0.5 keV 
by \citet{haberl94} using ROSAT data. Moreover, it was found that 
the temperature of the soft component depends on the contribution of the 
highly absorbed hard power-law component from the neutron star. 
According to \citet{Hickox04}, the observed
soft excess is due to emission by photoionized or collisionally
heated diffuse gas or thermal emission from the surface of the
neutron star.

Above 20 keV, cyclotron resonance scattering features (CRSFs) 
between 25 and 32 keV \citep{makishima92,choi96,krey02} and 
at $\sim$ 55 keV \citep{kendziorra92,orlandini98,labarbera03,attie04} have
been reported, although the interpretation of the 25 keV feature
is still sometimes debated \citep{orlandini06}.

Observations of the system during the eclipse of the X-ray pulsar with 
\textit{Tenma}, \textit{ASCA}, and \textit{Chandra} have revealed various 
fluorescent lines in addition to highly ionized lines and radiative recombination continua 
that imply the existence of optically thick and clumped matter in addition
to warm ionized plasma \citep{sato86,nagase94,sako99,schulz02}.

The structure of the wind of the system was investigated above 20~keV by 
\citet{fuerst2010}, who suggests that a mixture of a clumpy wind, shocks, 
and turbulence could explain the measured mass distribution. These authors 
estimate a mass of the wind clumps of the order of $m_{cl}=10^{21}$ g for 
the giant flares. Furthermore, in this work a log-normal distribution of 
the brightness was found along the orbit.

\citet{odaka13} report on \textit{Suzaku} observations of
\object{Vela X-1} covering the orbital phase interval 0.20--0.39 for
phase zero at $T_{90}$. They found strong variability with $L_\mathrm{X}$
(0.1--100~keV) ranging from $\sim 0.5$ to $\sim 9.5\times 10^{36}$~erg~s$^{-1}$
and variability time scales of 1--10~ks. Assuming a clumpy wind as the source
of the variations, the observed variability indicated clump radii of $(2-20)\times 10^{10}$~cm 
for a relative wind velocity of 400~km~s$^{-1}$. For a bright 
($L_\mathrm{X}$=$10^{37}$~erg~s$^{-1}$), short (1~ks) flare this would lead to 
a clump mass of $4\times10^{19}$~g and an overdensity
of two orders of magnitude compared to a smooth wind model, while longer,
less intense flares would be explained naturally by density variations of a
factor of  a few.

To study the structure of the extended atmosphere and the wind in
the Vela~X-1 system, a 123 ks {\it XMM-Newton} \citep{jansen01} observation was performed
immediately following eclipse egress. The main goal was a precise 
measurement of the evolution of the absorption column in this part of the orbit. From historical
observations \citep{sato86,lewis92}, one expects a steep decrease in the 
absorption measure $N_{\rm H}$ in this part of the orbit as the neutron star 
emerges from behind the companion's extended atmosphere. 

The remainder of the paper is structured as follows. In Sect.~\ref{obs},
the data and the software used are described. Section ~\ref{lc} describes
the light curves of the observation. Spectral analysis results are
presented in Sect.~\ref{spe}. Finally a discussion is given in Sect.~\ref{disc}
and a summary and conclusions in Sect.~\ref{sum}.


\section{Observations and data reduction}\label{obs}

XMM-Newton observed Vela X-1 on 2006 May 25-26
(MJD\,53880.452 to MJD\,53881.876) using the European Photon Imaging Camera (\textit{EPIC})
and the Reflection Grating Spectrometers (\textit{RGS}) during all of revolution 
1183 of the satellite. For this observation (ObsID~0406430201), the EPIC-MOS1 
(EPIC-Metal Oxide Semi-conductor) camera was disabled to maximize the 
available telemetry rate for the EPIC-pn CCD camera, which was set up in timing mode with 
the thin filter. In this mode, the camera provides timing information for each event 
with a resolution of up to 1.5\,ms, as well as the standard energy resolution 
of about $E/\Delta E = 40$.

Avoiding times close to the Earth's radiation belts, we used data between 
MJD\,53880.613 and MJD\,53881.768 (about 100 ks), corresponding to
the orbital phase interval 0.132 to 0.270 according to the ephemeris of \citet{krey08} 
(see Table~\ref{tab:ephem}), where phase zero is defined as $T_\mathrm{90}$. 
When comparing our findings with other results, one should note that  
sometimes $T_\mathrm{90}$ and sometimes the mid eclipse time 
$T_\mathrm{ecl}$ are used in the literature to define zero phase, where $T_\mathrm{ecl}$ is
later than $T_\mathrm{90}$ by 0.226~d \citep{krey08}. Using $T_\mathrm{ecl}$ as reference
would shift the orbital phase of our data set by 0.0252 to earlier phases.

In this paper we focus on the EPIC-pn data and its evolution along the orbital
motion of the neutron star around the supergiant companion. The RGS
data will be considered in later work. The data reduction was performed using 
SASv11.0 and CCFs as of 2012 May starting from ODF level running rgsproc and epproc. For EPIC-pn,
we applied a rate-dependent CTI correction, and to the time column a
barycentric correction, as well as a correction for the Vela~X-1 binary
system using the ephemeris as of \citet{Bildsten1997} and
\citet{krey08}. 


   \begin{figure}
   \centering
\includegraphics[width=0.95\linewidth]{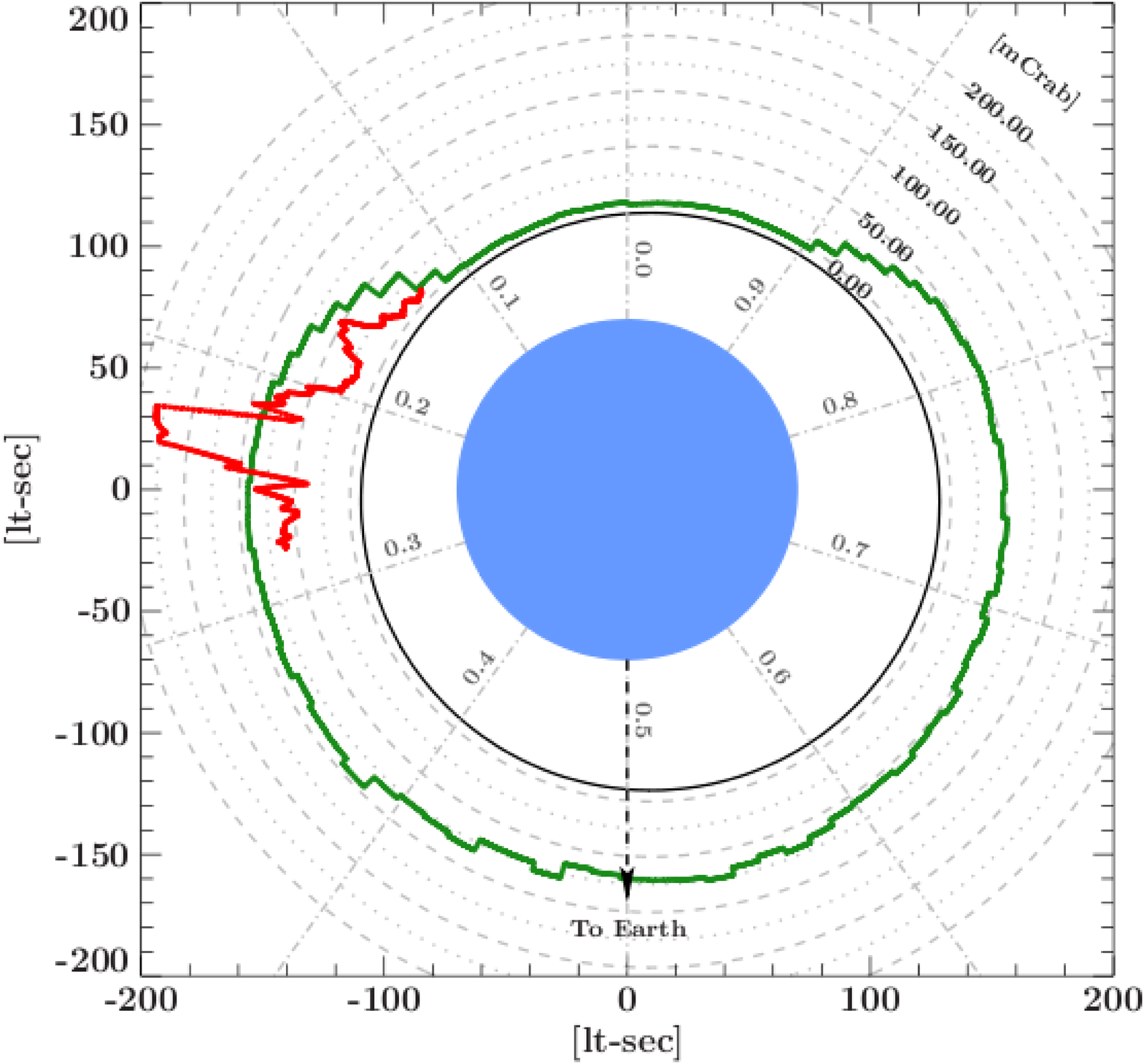}
   \caption{Orbit sketch of Vela X-1. The XMM-Newton 0.5-12 keV observed count rate
is shown in the red curve and compared with the average light curve of the source over
years of 1.5-12 keV RXTE/ASM observations (green curve).}
      \label{orbit}
   \end{figure}

\begin{table}[h]
  \centering
  \caption{Ephemeris data used for timing calculations taken
           from (1) \citet{Bildsten1997} and (2) \citet{krey08}.
           \label{tab:ephem}}
  \vspace{3mm}
  \begin{tabular}{lcc} \hline \hline
    $T_{\rm 90}$  & 52974.001 $\pm$ 0.012 MJD & (2) \\
    $P_{\rm orb}$ & 8.96436 $\pm$ 0.00003 d & (2) \\
    $a \sin i$    & 113.89 lt-s             & (1) \\
    $e$                & 0.0898 $\pm$ 0.0012       & (1) \\
    $\omega $     & 152.6 $\pm$ 0.9 $^{\circ}$      & (2) \\ \hline
   \end{tabular}
\end{table}

\subsection{Spectral extraction}

Due to the high count rate ($>$2500 cts/s) during the flare that occurred
as well as to the hard spectral emission
characteristics of the source, we had to cope with two instrumental
effects before extracting spectra: pile-up and the improvement of the
timing mode charge transfer inefficiency due the high amount of
shifted charge (rate-dependent CTI). The first can cause a change in
spectral parameters, e.g. a hardening of the continuum with increasing
levels of pile-up, the latter a general gain/offset shift of the
complete spectrum. We mitigated the effects of pile-up by
considering the pattern distribution (\emph{epatplot}), the maximum
count rate ($<$~800~cts/s),\footnote{according to the XMM-Newton Users
Handbook} and the stability of the spectral parameters dependent on
the columns used to extract the spectra. We found that the three columns
covering the central part of the PSF are significantly
affected by pile-up, and from spectral stability analysis we got
indications that their adjacent columns might still show slight and
close to insignificant pile-up effects. 

Restricting the analysis to the flaring period, we extracted
spectra taken from the full PSF and from regions ignoring
3, 5, 7, 9, and 11 columns of the PSF centre (named PSF-\# afterward)
and compared their modelling of the instrumental gold edge and the relative
position of the iron line, as well as general slope of the continuum. 

Without correction we found a shift of the gold edge of 93~eV keeping
the PSF centre, and we needed to exclude the central 11 columns
(PSF-11) to be consistent with no shift of the gold edge. The Fe 
line-energy shift continuously decreased with increasing number of 
ignored columns by 132~eV from full PSF to PSF-11.

The SAS-task \emph{epfast} is designed to correct for
these rate-dependent CTI effects seen in EPIC-pn fast modes (but not
for pile-up effects). Applying the epfast correction (using the corresponding public CCF-file
EPN\_CTI\_0023.CCF), we still found shifts at the gold edge of 13~eV
using PSF-7, 12~eV using PSF-9, and 7~eV using PSF-11, which is
consistent with the systematic uncertainties of the
correction\footnote{\tiny{XMM\_CCF\_REL\_256 on \\
http://xmm2.esac.esa.int/external/xmm\_sw\_cal/calib/rel\_notes/index.shtml}}.  
However, the Fe line energy was consistent for all spectra of
PSF-5 to PSF-11, but at an energy of about 40~eV higher compared to
the uncorrected PSF-11 value. 

We calculated the spectral shifts that depends on the amount of charge
per column for the flare period, and with support of the XMM-Newton SOC
we added the results to the calibration of the rate dependent CTI,
creating a non-public EPN\_CTI\_0024.CCF. Using this special
CCF, no more shifts at the gold edge were seen for all spectra of
PSF-5 to PSF-11, and the Fe line energy was consistent with the
uncorrected PSF-11 value. Also, the continuum slopes were consistent
within errors, with PSF-5 slightly harder than PSF-7 to
PSF-11. Finally we decided to exclude the innermost seven columns from the
spectral extraction and calculated a corresponding arf response
file. A comparison of the full PSF spectrum and ARF with our
PSF-7 spectrum and ARF for the pre-flare period yielded a systematic error
of about 3\% for the PSF-7 flux normalizations. For consistency, we use the 
PSF-7 region for all epochs of our data set. 

\begin{figure}[h]
 \centering
\includegraphics[angle=-90,width=0.99\linewidth]{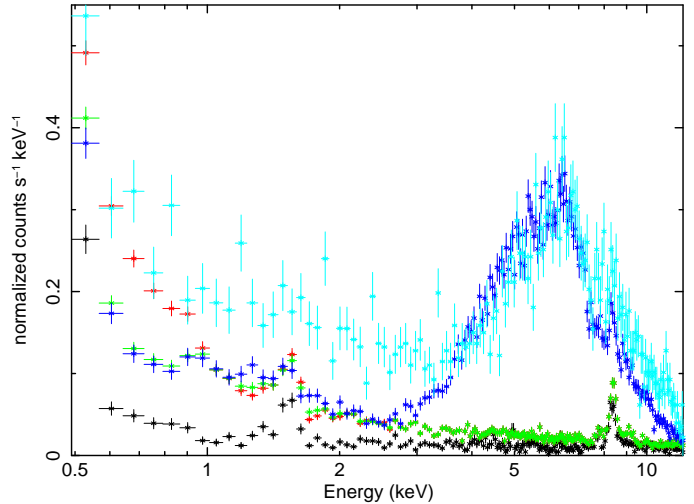}
\caption{Different timing mode backgrounds. Black: closed filter; 
 cyan: this Vela~X-1 (Obs.ID 0406430201, columns 3-12);  navy blue:
 Vela~X-1 (Obs.ID 0111030101, columns 3-12); red: empty sky
 (Obs.ID~0406620201 columns 16-33,41-58); green: final merged background.}
\label{fig:backgrounds} 
\end{figure}


\subsection{Background generation}

Looking into the columns closest to the timing mode window border, it
was obvious that the observation does not provide a local background
area, since a) the light curve extracted from these columns follows the
overall light curve of the source and b) its spectrum differs
significantly from what is expected to be an empty sky emission,
resembling the source spectrum. A background extracted from an earlier
timing mode observation of Vela~X-1 (Obs.ID 0111030101) showed
a similar pattern but one restricted to higher energies. We therefore
searched a timing mode "empty sky" and found Obs.ID~0406620201, where
no source is visible in the timing mode window. Excluding again the
innermost seven columns (identical PSF-7 extraction region), we
extracted a spectrum of Obs.ID~0406620201 and 
used it as an empty sky background. Since the local background of the
earlier Vela~X-1 observation (Obs.ID 0111030101) shows a lower
background level than this "empty sky", we merged these two
backgrounds for energies below $\sim$ 2.5~keV, scaled according to exposure
times and surfaces, to address the differences of absorption columns
of about factor 100-1000.  




   \begin{figure*}
   \centering
\includegraphics[angle=90,width=1.98\columnwidth]{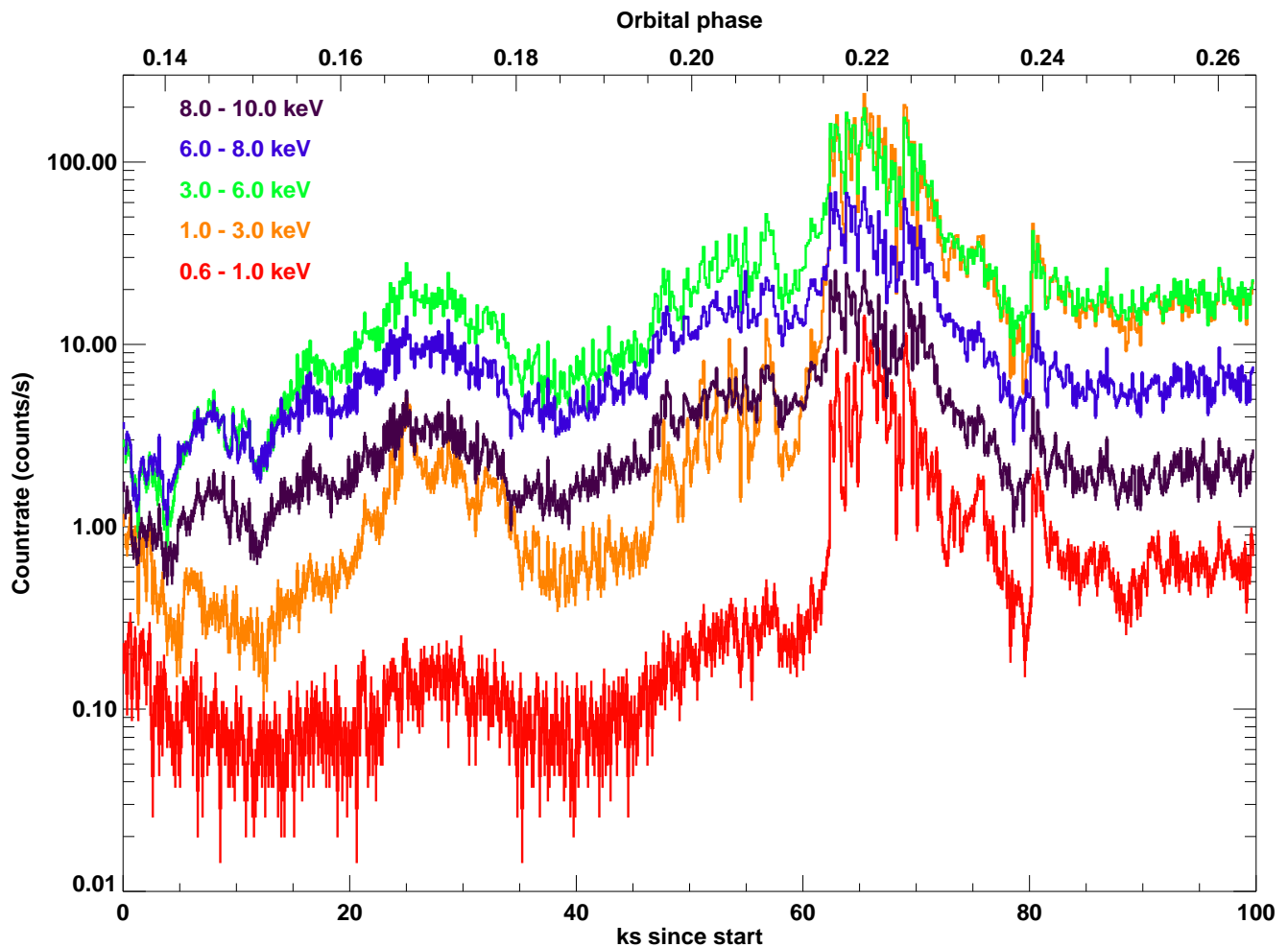}
   \caption{Light curve observed during the XMM-Newton Observation with the
  EPIC-PN camera with a binning time of 141s ($\sim$ half a pulse period). 
  The bands shown are 0.6--1, 1--3,  3--6, 6--8 and 8--10~keV. The count rate is 
  plotted on a logarithmic scale to emphasize the variability in the different energy bands.}
\label{LIGHTCURVE_COMPARISON}             
   \end{figure*}

\begin{figure*}
\begin{center}
\includegraphics[width=1.85\columnwidth]{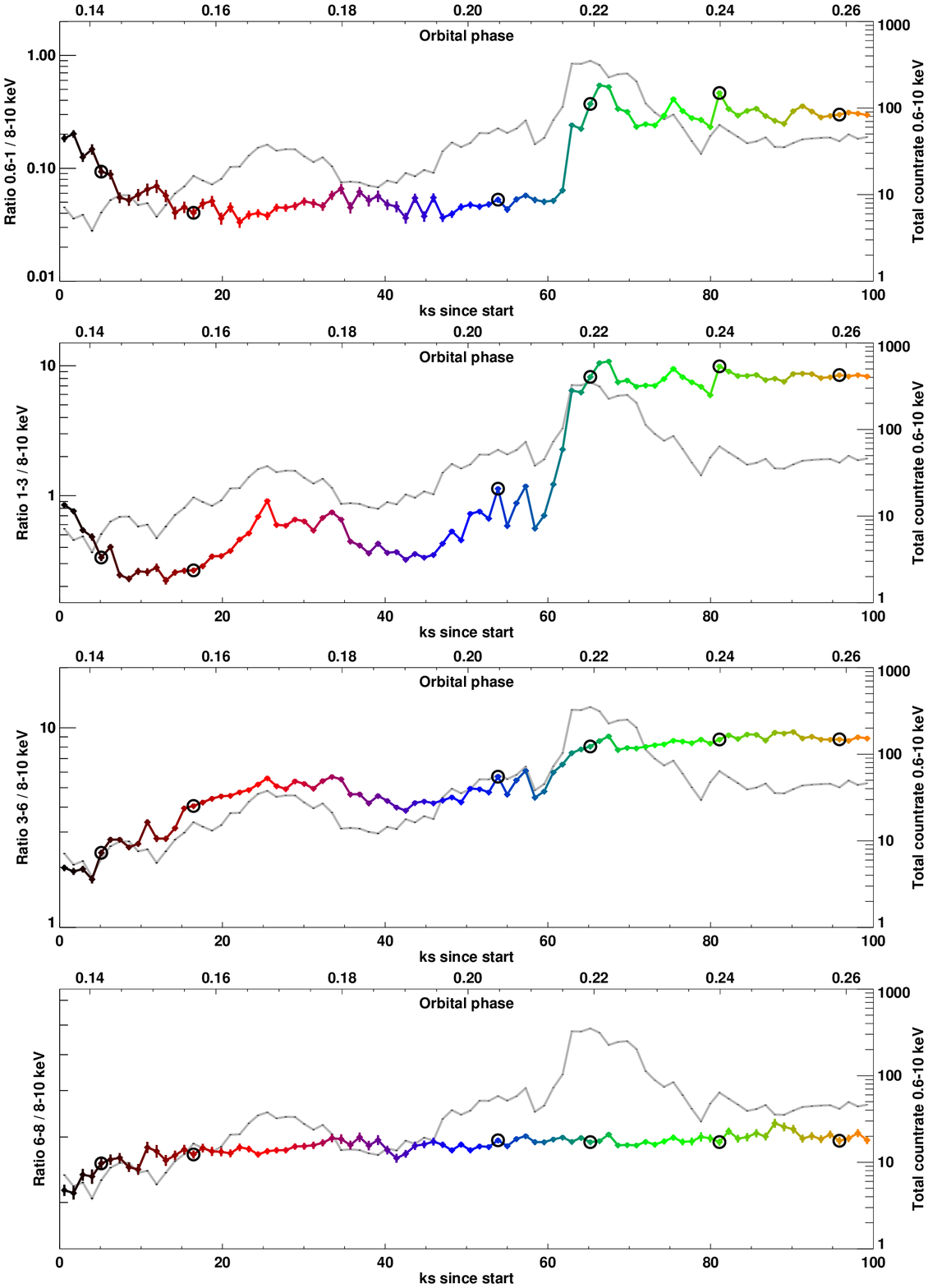}
\caption{Ratios between count rates in different energy bands to that of the 8--10~keV band.
From top left to bottom right: 0.6--1~keV, 1--3~keV, 3--6~keV, and 6--8~keV. The
time binning is 1128~s ($\sim$4 pulse periods), matching the spectra used in the
time-resolved spectral analysis. In the background, the total flux evolution is
plotted for comparison with the light curves. Circles mark the data points corresponding
to the spectra shown in Fig.~\protect{\ref{pn_phase_ex}}}
  \label{LIGHTCURVE_RATIOS}
\end{center}
\end{figure*}


\section{Energy-resolved light curves and flux ratios}\label{lc}

During the course of the observation, Vela~X-1 displays strong variations in flux,
both overall and in the relative flux in different energy bands. These variations
are driven by a mix of the intrinsic variability of the X-ray source and the
modulation of the X-ray flux by absorption and scattering within the system.
These variations happen on all time scales studied, from tens of kilo-seconds down
to fractions of the pulse period. We have not undertaken a detailed analysis of
the timing behaviour on time scales below the pulse period. A corresponding
study is under way and will be published separately. In this work a mostly 
stable pulse period of  283.389$\pm$0.004\,s is found. This pulse period 
was measured using a phase-connected technique over 10 ks intervals. The mean 
photon arrival times for each interval were then corrected by an 
empirical model and fit using a polynomial function.

To describe the overall evolution, we have generated light curves in
several energy bands using the approach and corrections as described 
in Sect.~\ref{obs}. The bands used are 0.6--1, 1--3,
3--6, 6--8, and, 8--10~keV, respectively. Figure~\ref{LIGHTCURVE_COMPARISON} shows 
the light curves in the different bands for comparison. The ratios between the count rates 
in the first four bands and the 8--10~keV band, which is least affected by the strong absorption variations,
are displayed in Fig.~\ref{LIGHTCURVE_RATIOS}. 

Based on the light curves and their relative rates, several distinct 
intervals can be defined on the time scale of tens of kilo-seconds, as explained
in the following, where we used $T_{\mathrm{obs}}$ to denote the time since the 
start of the observation (MJD\,53880.61305). During the first $\approx10$~ks the 
total flux varies moderately, but as Fig.~\ref{LIGHTCURVE_RATIOS} shows, there are opposing trends in the
flux ratios below and above 3~keV. After this interval, the ratio \mbox{6--8/8--10~keV} 
remains nearly constant despite large flux variations, indicative of a stable 
spectral shape for the unabsorbed source. A broad peak at all energies is evident 
between $T_{\mathrm{obs}}\approx10$~ks and $T_{\mathrm{obs}}\approx40$~ks, which is most 
pronounced in the 1--3~keV band.

From $T_{\mathrm{obs}}\approx40$~ks to $T_{\mathrm{obs}}\approx60$~ks the source 
flux starts to rise in all energy bands. On top of the general rise, there is 
flaring visible on time scales of ks between $T_{\mathrm{obs}}\approx45$~ks and 
$T_{\mathrm{obs}}\approx55$~ks, again most pronounced in the 1--3~keV band. Around 
$T_{\mathrm{obs}}\approx60$~ks, the total source flux rises rapidly,
more than an order of magnitude within a few kilo-seconds. At energies $\ge$6~keV,
the flux actually changes only by a factor of a few, but the contribution
$\le$3~keV rises by several orders of magnitude, while the flux ratio
6--8~keV/8--10~keV remains stable.

Around the peak of the flare, from $T_{\mathrm{obs}}\approx62$~ks to $T_{\mathrm{obs}}\approx70$~ks,
the overall flux is strongly variable with a series of spikes and dips. 
These structures are highly correlated across all bands, but 
variations in the flux ratios are mainly visible
in the 0.6--1 and 1--3~keV bands, such that higher global flux is
mainly driven by higher contributions from the low energies. From $T_{\mathrm{obs}}\approx70$~ks to $T_{\mathrm{obs}}\approx78$~ks
the flare decays with the overall flux diminishing by an order of 
magnitude, similar in all bands. Around $T_{\mathrm{obs}}\approx80$~ks, a smaller, but still significant
flare is visible, where again any spectral variation is mainly visible
at lower energies.

Beyond $T_{\mathrm{obs}}\approx82$~ks, up to the end of the observation,
the source seems to settle. While the flux still varies randomly
in all bands, albeit less than during earlier phases, 
the ratios between different bands do not show a pronounced evolution.

While the short term variability appears to be mostly 
dominated by the intrinsic flux variations, it is important 
to note that the strong absorption in the circumstellar
wind has a marked effect on the overall observed 
count rates. Moreover, the dramatic change at the onset of
the flare seems to be driven by a significant variation in
the absorbing and reprocessing material between the source
and the observer. 



\begin{figure*}
\centering$
\begin{array}{lr}
\includegraphics[width=0.44\linewidth]{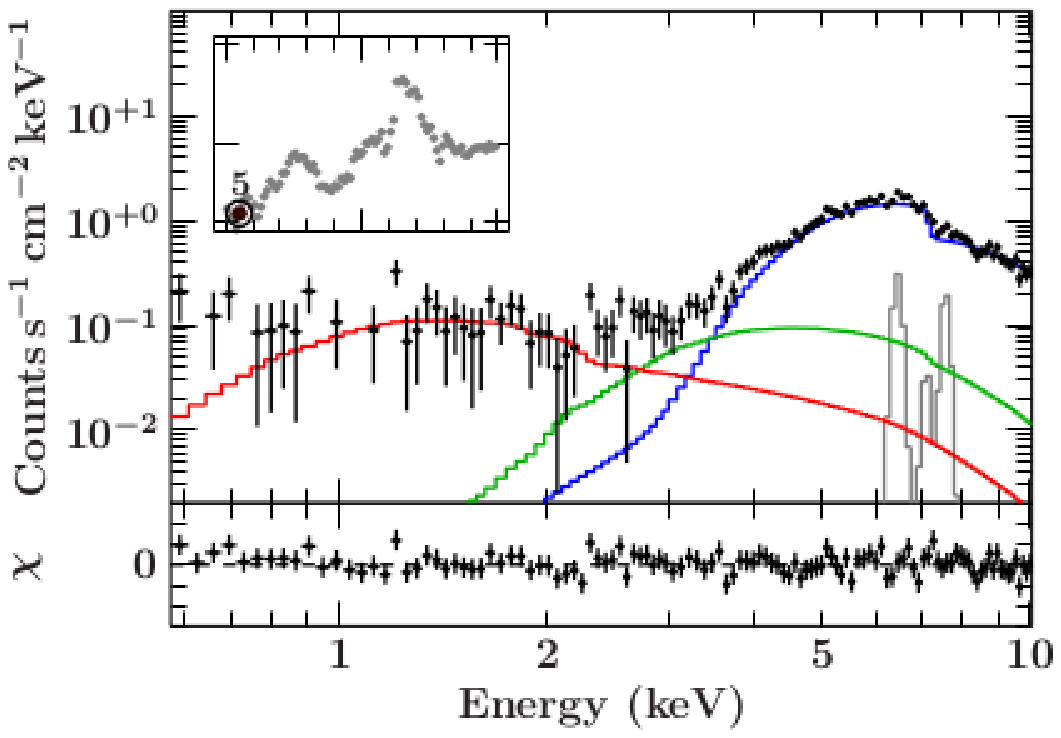} & \includegraphics[width=0.44\linewidth]{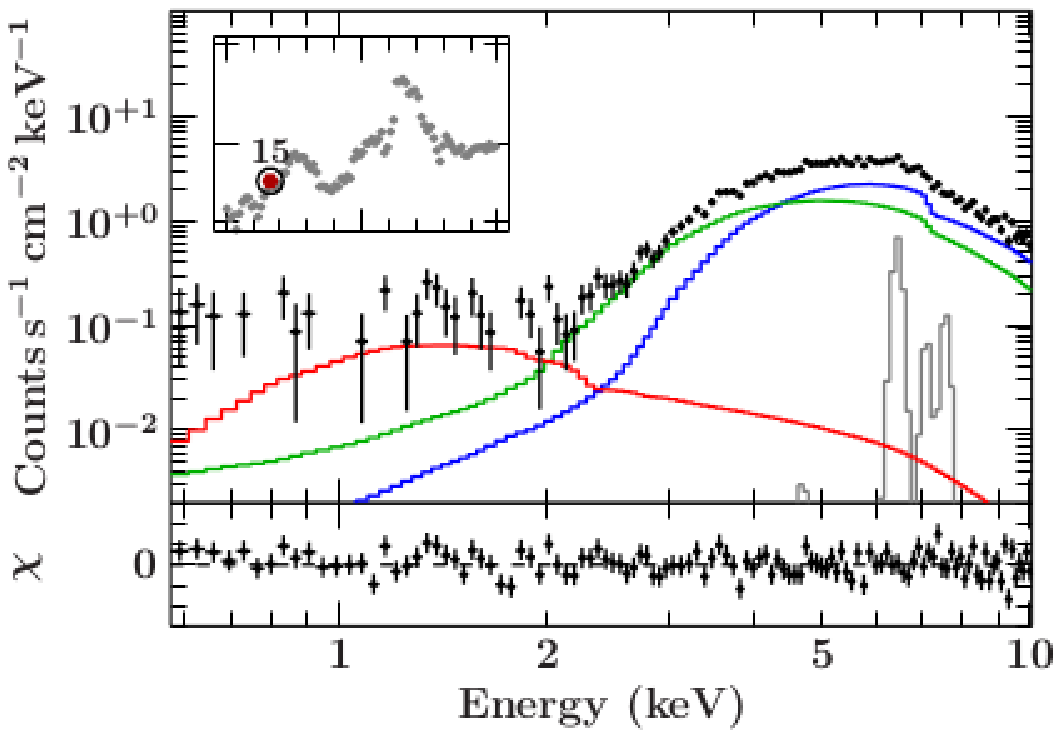}\\
\includegraphics[width=0.44\linewidth]{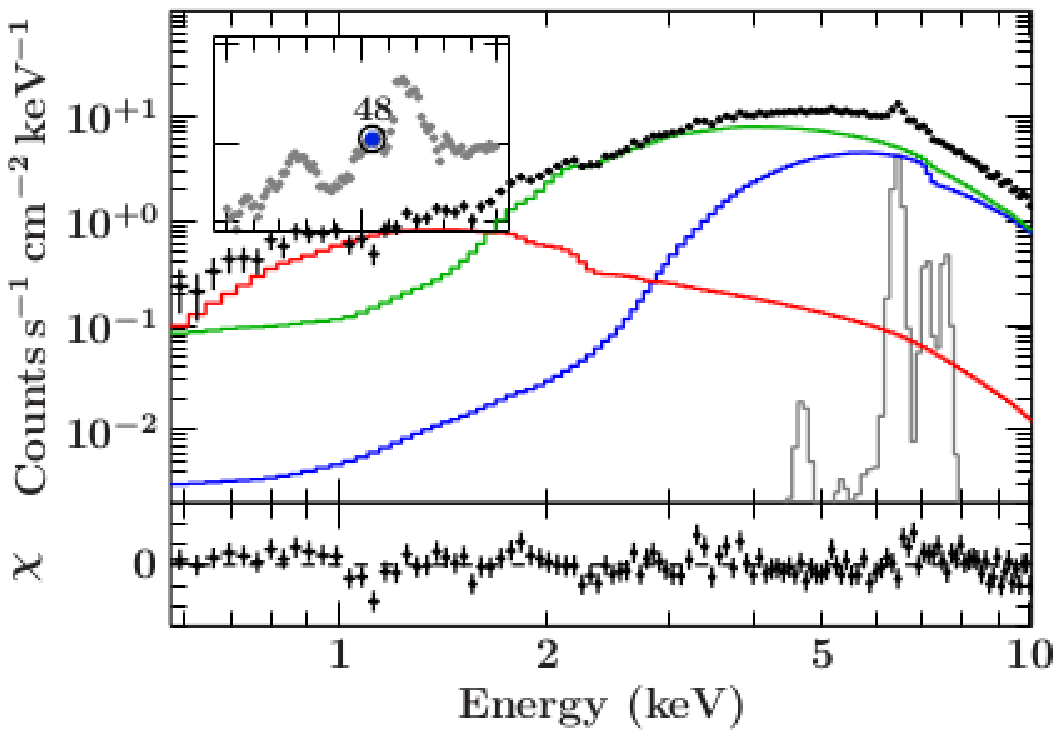} & \includegraphics[width=0.44\linewidth]{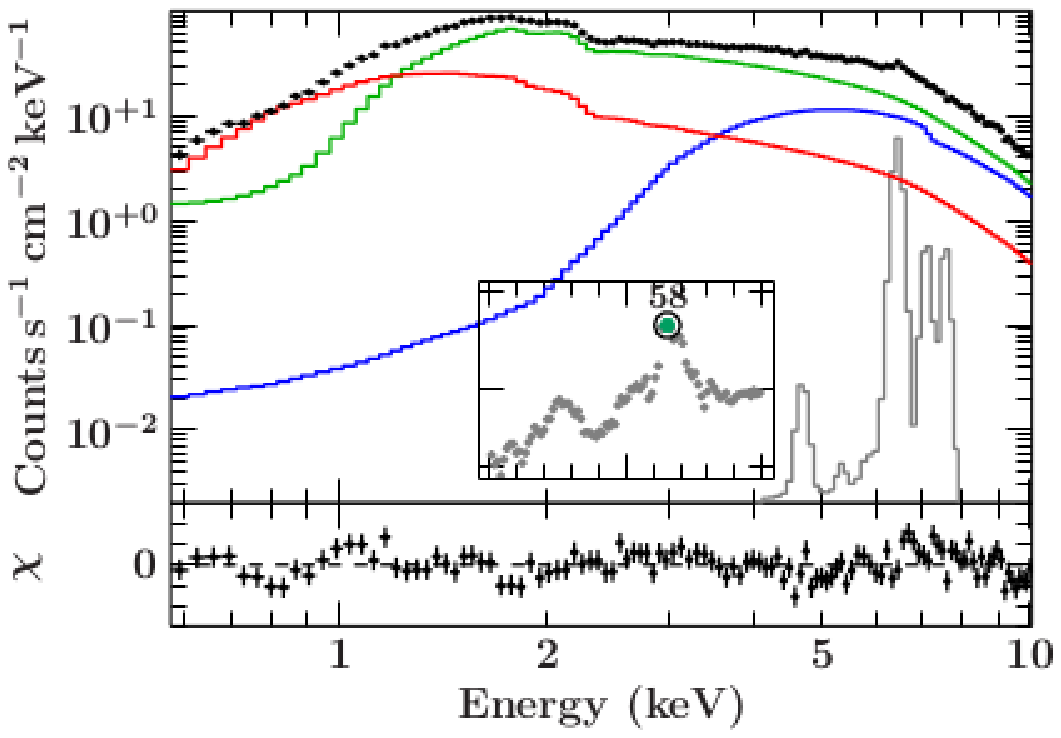}\\
\includegraphics[width=0.44\linewidth]{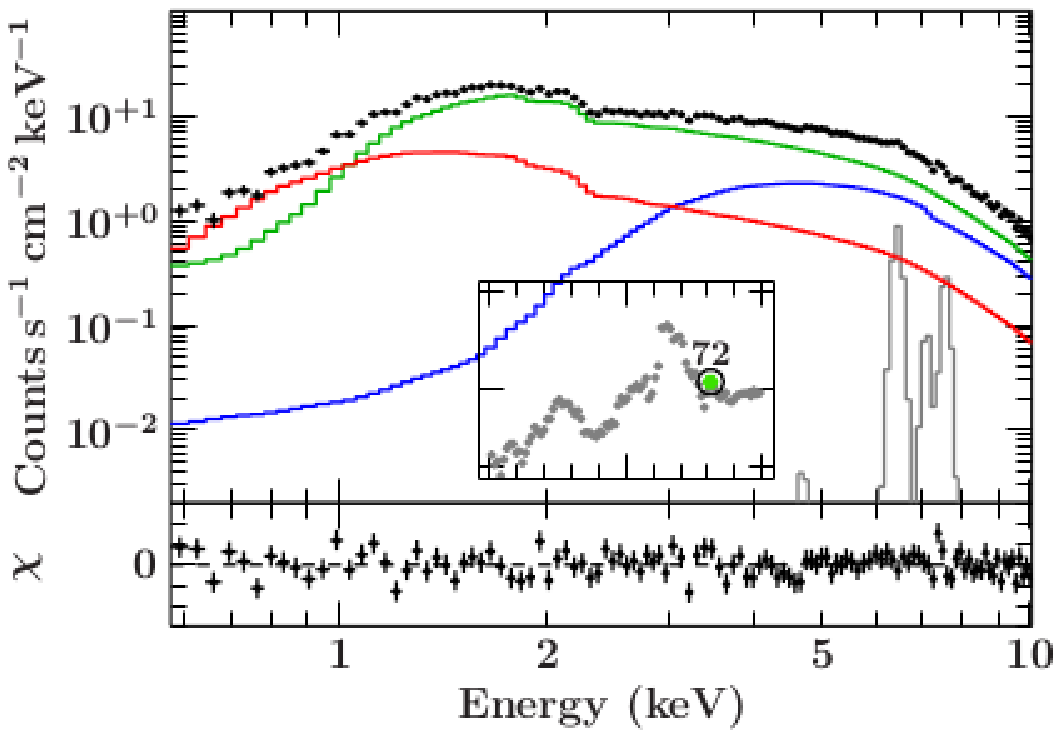} & \includegraphics[width=0.44\linewidth]{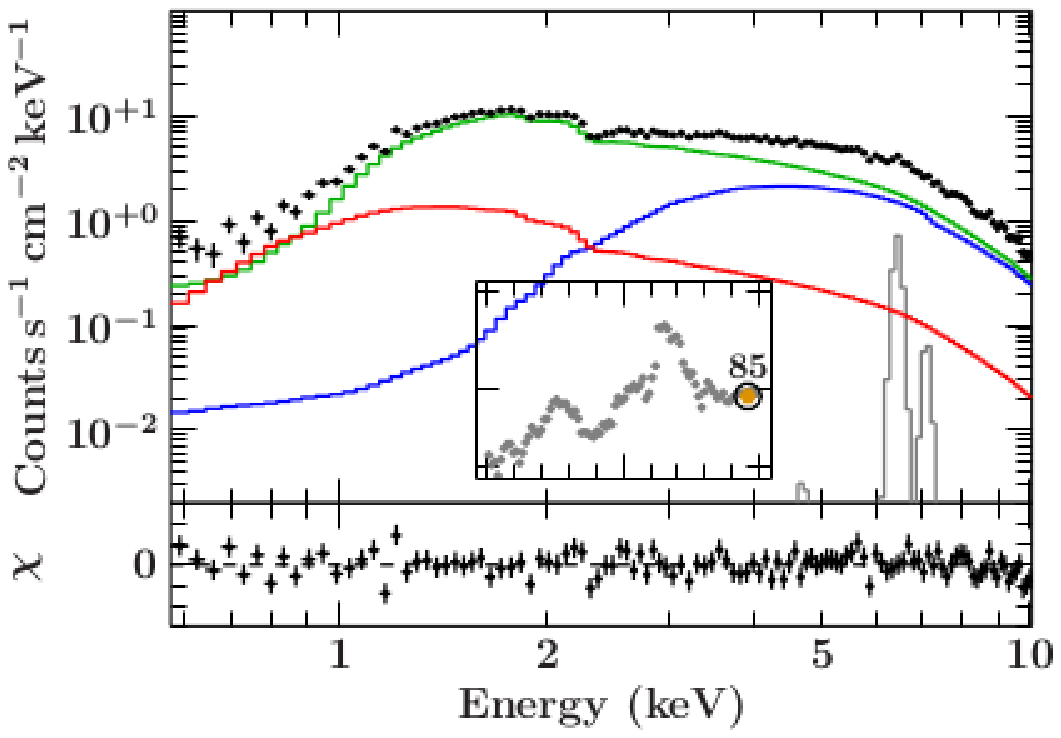}\\
\end{array}$
\caption{A representative sample of the 88 EPIC-pn spectra at different relevant times marked in the inset light curve 
plot. The different components of the model are shown: component 1 (blue), component 2 (green)
component 3 (red) and fluorescence lines  modelled as Gaussian functions (grey).}
\label{pn_phase_ex}
\end{figure*}


\section{Orbital phase-resolved spectroscopy}\label{spe}

To characterize the evolution of the stellar wind inhomogeneities
surrounding the neutron star, a study of the flux variations in combination with 
the changes in the absorption column is carried out, under the hypothesis that these variations 
are proportional to the local $\dot{M}$. Therefore we have divided
our data in 88 EPIC-pn spectra of 1.1 ks exposure time each. The spectral time 
resolution was chosen as a compromise between the statistics of the 
data and the short term variations observed in the light curve analysis. Some 
of these spectra are shown in Figure~\ref{pn_phase_ex} at different relevant periods.

Since a fully physical model of the X-ray production mechanism in accreting X-ray pulsars 
does not yet exist owing to the complexity of the physical problem (significant progress has, however,
been made, see e.g. \citep{becker07}), an empirical model has to be used. After trying different 
models, we have found that there is only one continuum model that could fit 
all the spectra significantly well throughout the whole observation \textit{including the flare}. 
It consists of three absorbed power-law components with a unique photon index but 
different normalization factors for each component. Furthermore, the spectra of the source 
clearly show the presence of three emission lines corresponding to Fe K$_\alpha$, Fe K$_\beta$,
and Ni K$_\alpha$ fluorescence lines. These lines have been treated as being absorbed by the 
highest absorbed component and modelled by a Gaussian function even if their shapes 
are not perfectly Gaussian. A full description of the model is given in equation~\ref{eq1}.

A similar continuum model has been used previously to characterize the spectra of several 
high-mass X-ray binary systems, such as 4U 1700-37 \citep{vanMeer05}, Cen X-3 \citep{Ebisawa96},
and 4U 1538-52 \citep{Rodes11}. Our model differs from those models in two aspects: 
(a) we used a more complex ISM absorption model (\textit{tbnew} in equation~\ref{eq1}) 
and (b) we applied a non-relativistic Compton scattering correction to the 
first and second spectral components because of their large
measured hydrogen column densities (\textit{cabs} in equation~\ref{eq1}).\\


\begin{eqnarray}\label{eq1}
F(E) &=&  \sum_{i=1}^{2} \left[tbnew_{i} \times \mbox{cabs}_{i}         
                         \times \mbox{Norm}_{i} \times   E^{-\Gamma}\right] \nonumber \\
&& +\, tbnew_{3}  \times \mbox{Norm}_{3}   \times   E^{-\Gamma} \nonumber \\
&& + \sum_{j=1}^{3} \left[tbnew_{1} \times \mbox{cabs}_{1} \times \mbox{Gaussian}_{j}\right]
\end{eqnarray}


In equation~\ref{eq1} the summation index \textit{i} refers to the three model 
components, and the index \textit{j} refers to the three emission lines components. 
$Norm_{i}$ are the normalization factors of each power law (photons keV$^{-1}$ cm$^{-1}$ s$^{-1}$ at 1 keV), and $\Gamma$ is the power law 
photon index. These components are further modified by two effects both arising in 
the material between the X-ray source and the observer 
with density column $N_{H}$, on one hand, the photoelectric absorption. 
For this purpose we use the model \textit{tbnew}, an updated version of the T\"ubingen-Boulder 
ISM absorption model \citep{Wilms00}.  The absorption cross sections 
are adapted from \citet{Verner96}, and the abundances are set to 
those of \citet{Wilms00}. On the other hand, the non-relativistic Compton scattering effect is taken
into account by using a \textit{cabs} model, which is described by


\begin{equation}
\mathrm{cabs}(E)=\exp(-N_H\sigma_{T}(E)),
\end{equation} where $\sigma_{T}(E)$ is the Thomson cross section, and N$_H$ is the equivalent 
hydrogen column that it is linked to its corresponding absorption law.



\begin{figure*}
\centering$
\begin{array}{lcr}
\includegraphics[width=0.25\linewidth]{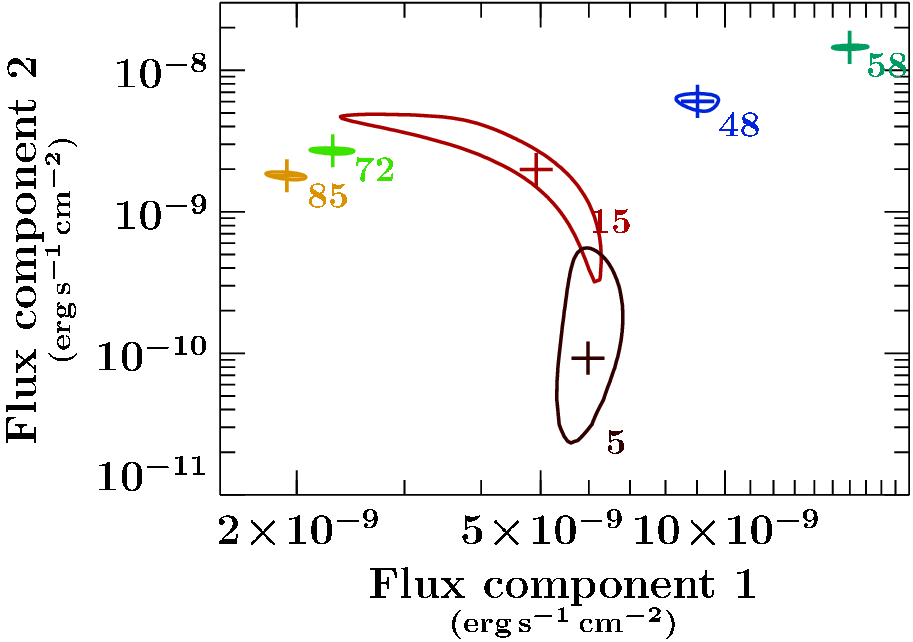} & \hspace*{1.5cm} & \includegraphics[width=0.25\linewidth]{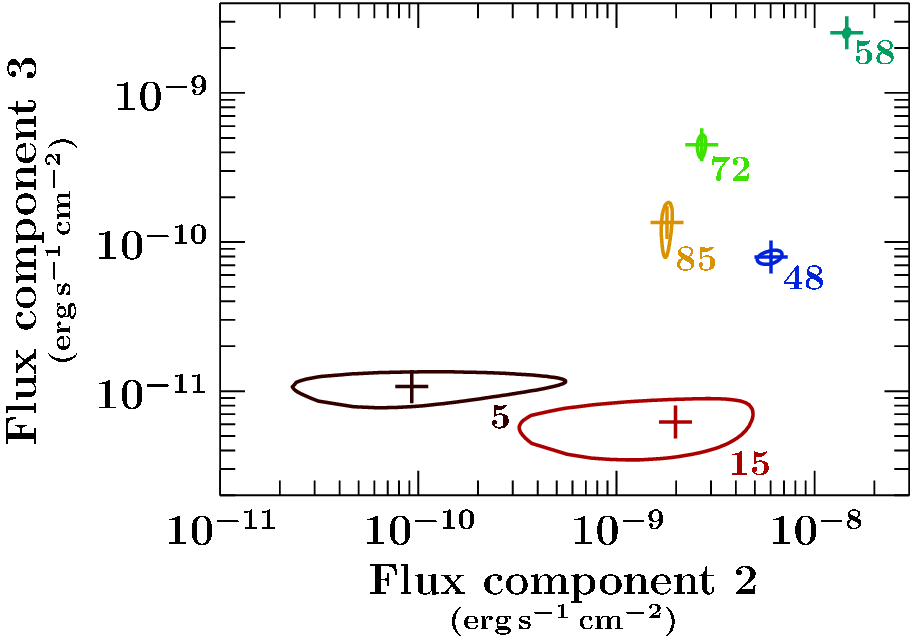}\\
\includegraphics[width=0.25\linewidth]{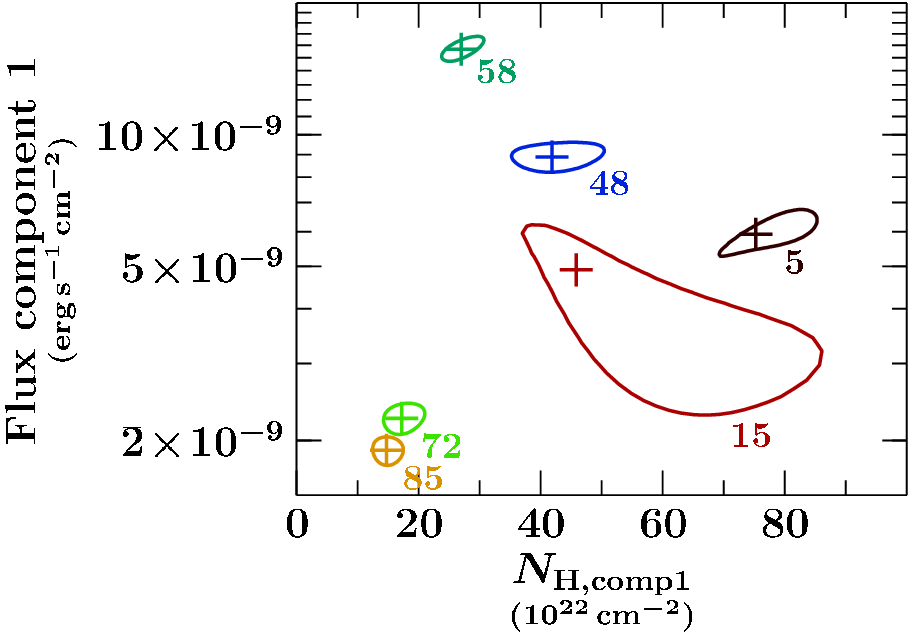} & \hspace*{1.5cm} & \includegraphics[width=0.25\linewidth]{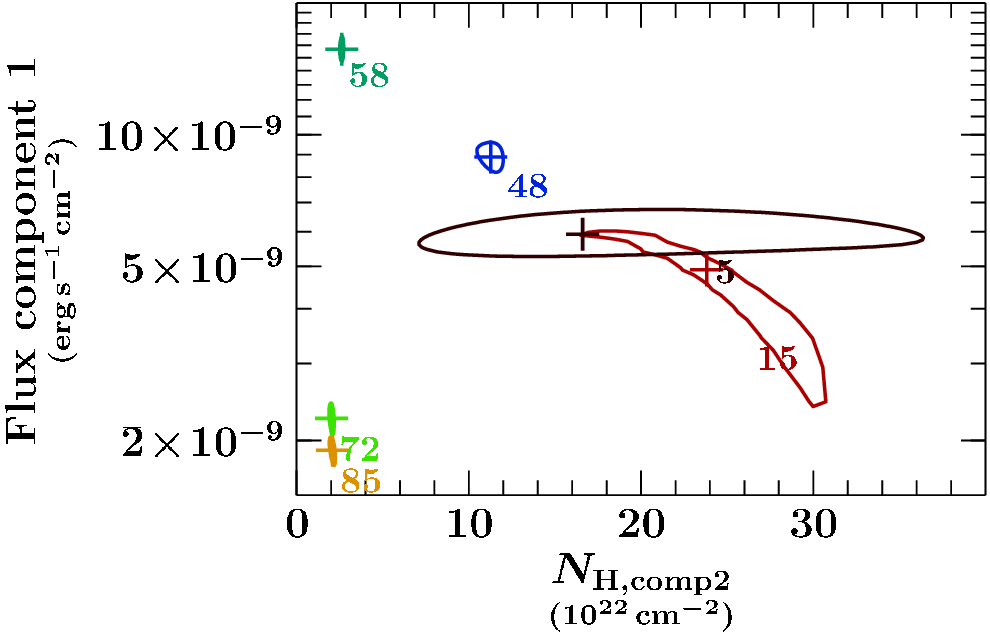} \\  
\includegraphics[width=0.25\linewidth]{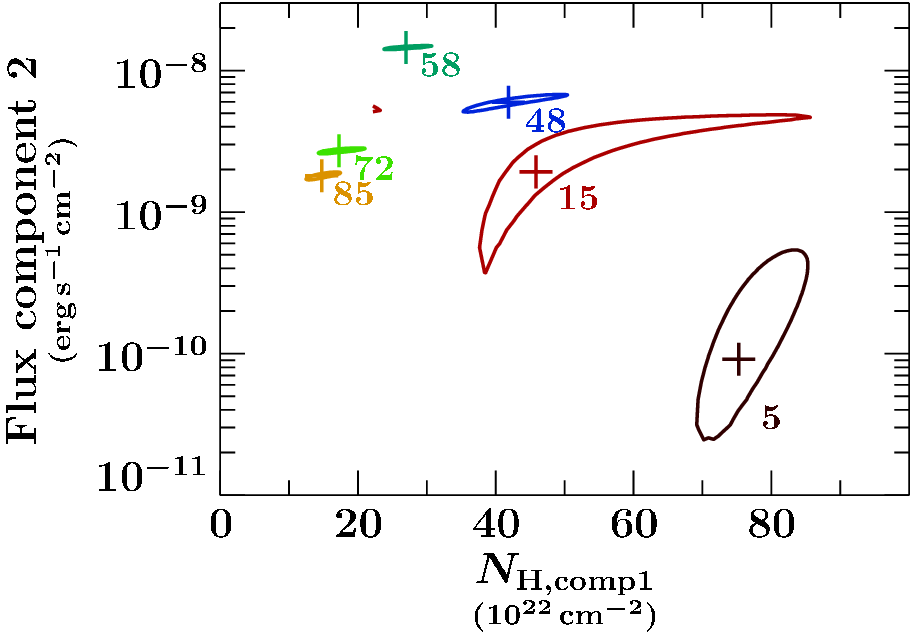} & \hspace*{1.5cm} & \includegraphics[width=0.25\linewidth]{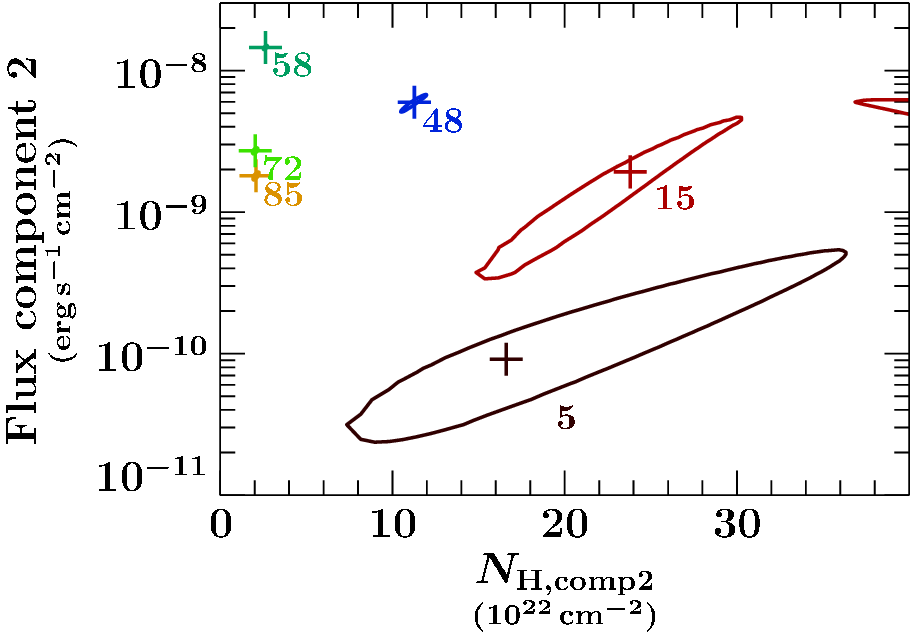} \\  
\includegraphics[width=0.25\linewidth]{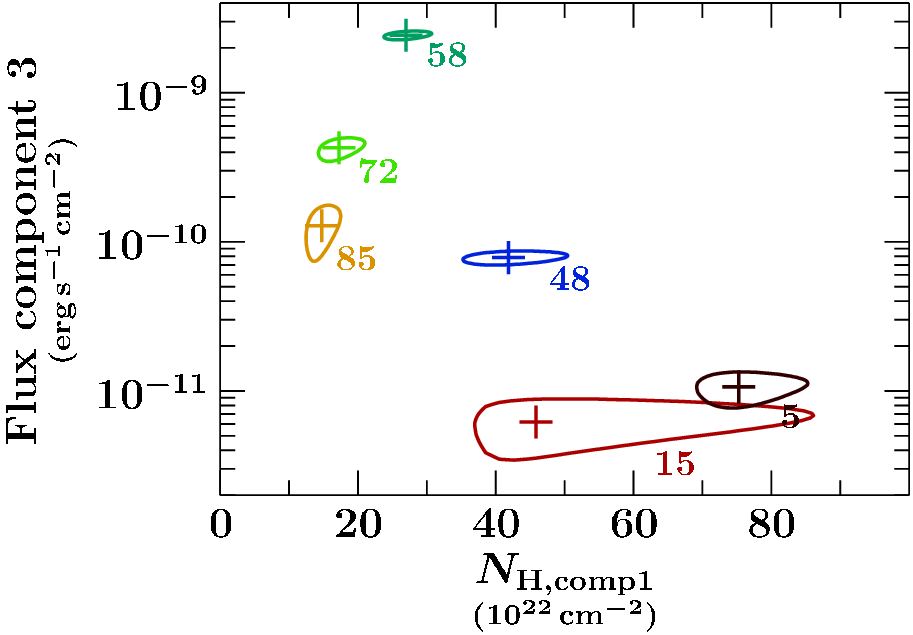} & \hspace*{1.5cm} & \includegraphics[width=0.25\linewidth]{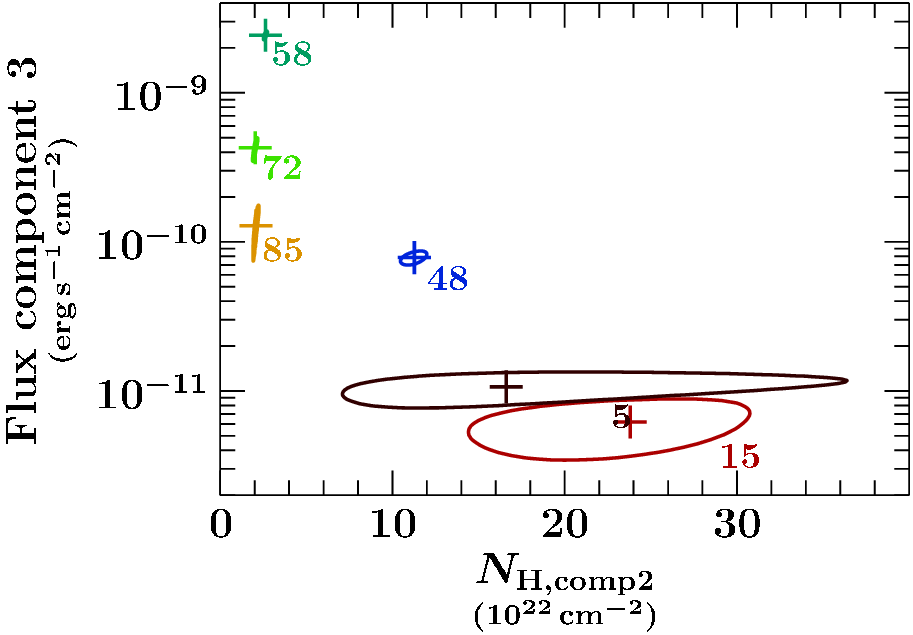} \\  
\includegraphics[width=0.25\linewidth]{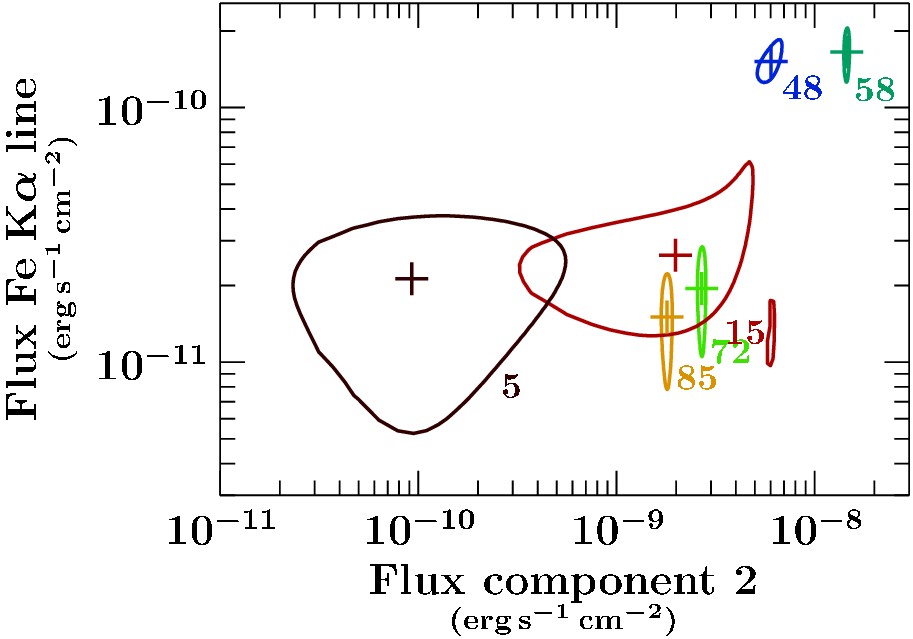} & \hspace*{1.5cm}  & \includegraphics[width=0.25\linewidth]{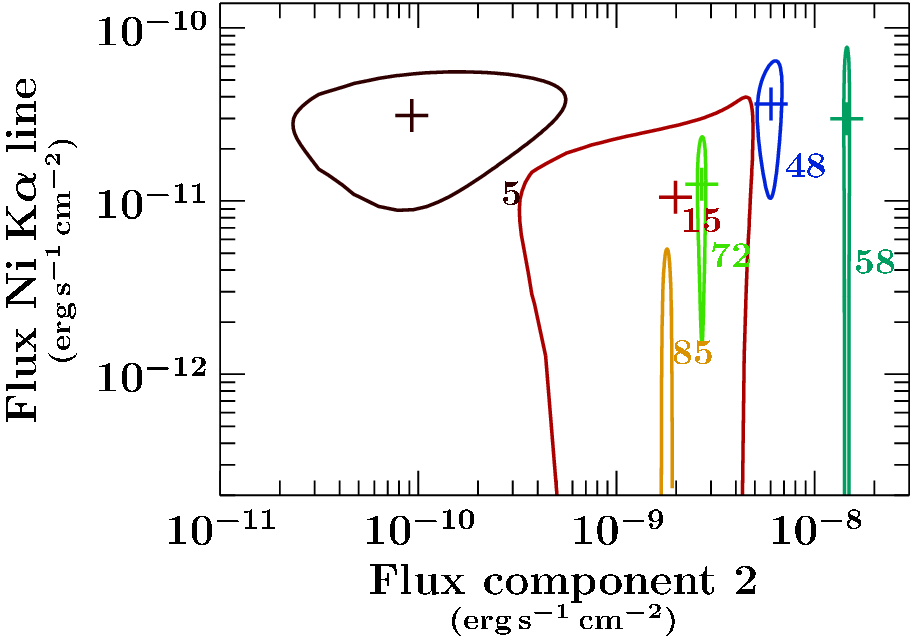} \\
\end{array}$
\caption{Contour maps (2$\sigma$ confidence limits for the two selected parameters) between the free spectral parameters for the six selected spectra shown
in Fig.~\ref{pn_phase_ex}. These maps demonstrate that while for some spectra the parameters cannot be constrained tightly, the observed changes of parameter values along the observation are real and significant.}
\label{cm}
\end{figure*}

In time- or phase-resolved spectroscopy, each spectrum is usually fitted individually
by minimizing the statistic, while here parameters are varied, and a minimum is sought across 
all 88 spectra simultaneously using the Interactive Spectral Interpretation System
(ISIS) software \citep{isis}. This novel approach allows finding a common continuum, as well as  
obtaining the best possible fit statistics and the best constraints of 
the relevant parameters. Figure~\ref{pn_phase_ex} show the spectra and 
fitted models for some representative time bins throughout the observation. 
The inspection of this plot already reveals that the most dramatic changes 
are produced in components~2 and 3.

In a first analysis we left the power law index $\Gamma$ free 
to vary among all 88 spectra. We found no significant differences among 
the best-fit values of this parameter for any of the 88 spectra. Consequently, 
in order to increase the precision of the other parameters, we assume a 
common value for $\Gamma$, i.e. a stable continuum slope for all 88 spectra. The best-fit 
value for this index is $\Gamma = 1.595\pm 0.010$ ($\chi^{2}$ = 14011 with 9765 d.o.f.). The 
absorption column $N^{(3)}_{\rm H}$ of the third component was found to be constant within errors. We 
therefore kept it fixed to the average value of ($0.75 \pm 0.03$) $\times$ 10$^{22}$ cm$^{-2}$. 
This value is in excellent agreement with the ISM absorption towards GP Vel, the optical counterpart 
of Vela X-1, derived from optical data. Finally, to pin down the fluorescence lines, 
we have constrained the Fe K$\beta$, as well as the Ni K$\alpha$ energies, with respect to the 
Fe K$\alpha$ centroid energy, taking theoretical calculations into account \citep{Kallman2004}. 
The energy of the Fe K$\alpha$ line (left free) turns out to be  6.435$\pm 0.001$ keV . The energy of the 
Fe K$\beta$ line is shifted by 0.65~keV, while its flux is considered
to be 13$\%$ of that of K$\alpha$. These energies are compatible with K shell fluorescence from ionized 
Fe up to a maximum of Fe\textsc{xviii} (\citet{Kallman2004}, Fig. 3). In the same way, the Ni K$\alpha$ centroid energy was shifted 
by 1.2 keV with respect to that of Fe K$\alpha$. These ratios were kept fixed during the fitting process. 

To estimate the possible dependence among the relevant parameters during the 
fits, we have computed confidence level contour maps for all 88 spectra. In Fig.~\ref{cm} 
we show contour maps for the same selected spectra as in Fig.~\ref{pn_phase_ex}.
As can be seen, the contours do not betray a systematic dependence, and 
therefore, the obtained parameters and their errors are considered reliable.


    \begin{figure*}
    \centering
    \includegraphics[width=0.85\linewidth]{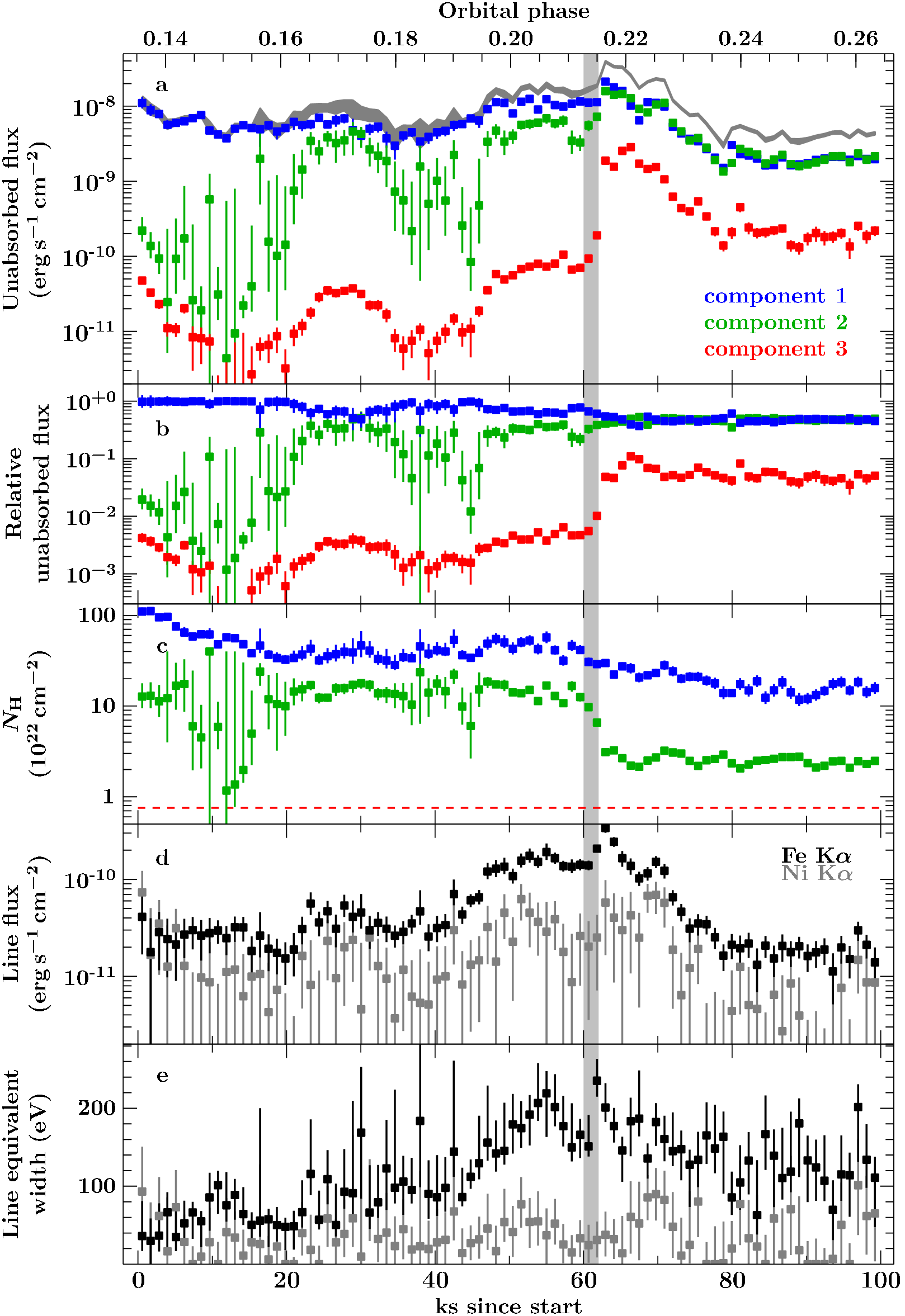}
    \caption{\textit{From top to the bottom:} Evolution of the model parameters versus time 
    and orbital phase. Error bars are 90\% confidence level. The dashed line in panel \textit{c} 
    indicates the constant $N_H$ of component~3. The grey column marks the rise of the flare and
    the grey curve in panel \textit(a) shows the overall unabsorbed flux of the spectral model.}
      \label{parm_evol}
    \end{figure*}

\subsection{Evolution of the parameters}

In Figure~\ref{parm_evol} we show the evolution of the relevant parameters of our model 
throughout time and orbital phase. Panel $a$ shows the unabsorbed flux evolution of
 each component and the overall unabsorbed flux (note the logarithmic ordinate scale). 
A clear progressive increase in the fluxes of components 2 and 3 can be seen, and meanwhile 
the flux of component 1 decreases slightly from the beginning to the end of 
the observation. On top of this we clearly observe two flares at $\phi\approx 0.17$ 
and 0.22 (a giant flare), respectively. A vertical grey line marks the rise of 
the bright flare. 
 
\begin{figure*}
\centering
\includegraphics[width=1.8\columnwidth]{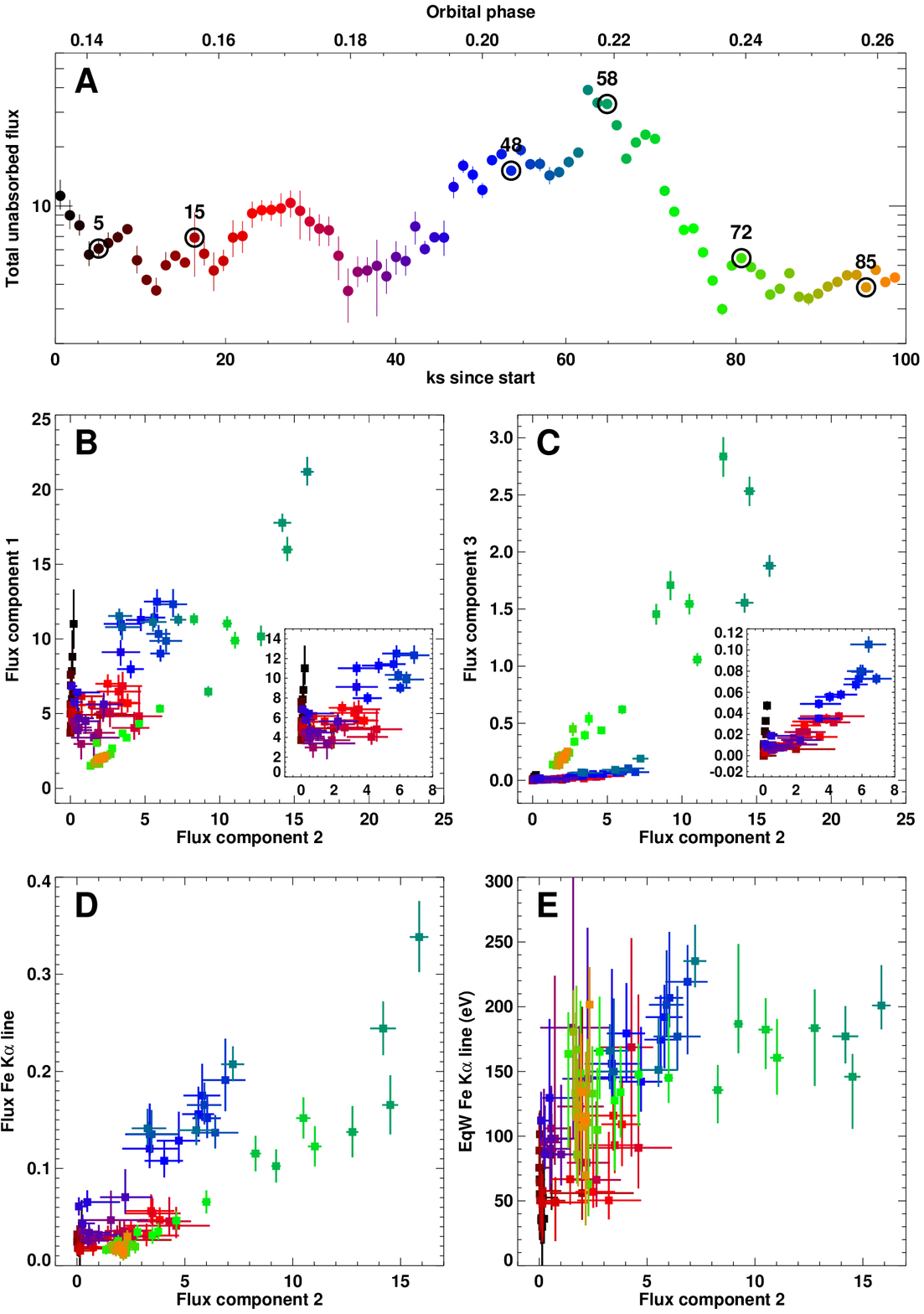}
\caption{\textit{A}: 0.6-10 keV unabsorbed flux light curve with a
binning time of 1.1~ks.
Marked circles indicate the data points corresponding to the spectra
shown in Figs.~\protect{\ref{pn_phase_ex}} and ~\protect{\ref{cm}}.
\textit{B}: Flux of the first component versus that of the second component. The inset shows an 
enlarged view of the first 58~ks, before the onset of the giant flare.
\textit{C}: Flux of the third component versus that of the second component with an inset 
equivalent to that in \textit{B}.
\textit{D}: Fe K$\alpha$ fluorescence line flux areas against normalization factor of second component.
\textit{E}: Fe K$\alpha$ equivalent width against the normalization factor of the second component. 
All fluxes are in units of $10^{-9}$ ergs cm$^{-2}$ s$^{-1}$.} 
  \label{corr}
\end{figure*}

In panel $b$ we show the relative unabsorbed flux evolution of 
each power law against the average flux of the first component. 
Several characteristics can be pointed out here. On 
one hand, as stated before, the most dramatic changes are 
seen in components 2 (scattered) and 3 (low energy). During the bright flare, 
component 3 increases by two orders of magnitude, while component 2 increases one order 
of magnitude.  These two components are very well correlated throughout the whole 
observation (except at the maximum of the bright flare). These can be clearly seen 
in panel $C$ of Fig.~\ref{corr}. In contrast, component 1 (direct) increases 
only by a factor of 2 during the flare. On the other hand, while components 
2 and 3 show an overall increase in flux throughout the observation, 
following the trend of the light curve (top panel), component~1 
shows an overall decrease in flux 
from $1\times 10^{-8}$ erg s$^{-1}$ cm$^{-2}$ to $0.2\times 10^{-8}$ erg s$^{-1}$ cm$^{-2}$. 
After the flare, the flux of components ~1 and 2 become identical within the 
errors, and in fact, the flux of all three components become very well correlated. 
This is clearly seen in panels $B$ and $C$ of Fig.~\ref{corr}.

In panel $c$ of Fig.~\ref{parm_evol} we show the evolution of the column density 
of the absorbing material for components 1 and 2 in units of $10^{22}$ cm$^{-2}$. 
The $N_{H}^{(3)}$ for component~3 remains constant throughout the observation 
at $\sim 0.75\times 10^{22}$ cm$^{-2}$, which is entirely compatible with the 
ISM value \citep{vanGenderen81}. As can be seen, the absorption of component~1 (direct) decreases 
from values of  $\geqslant 10^{24}$ cm$^{-2}$ at the beginning of the eclipse 
egress down to $\sim 1.8\times 10^{23}$ cm$^{-2}$ at quadrature. This reflects
the decreasing amount of stellar wind material in the line of sight 
towards the neutron star as it emerges from eclipse. An enhancement is present, 
however, between orbital phases 0.19 and 0.23. Remarkably, there is no dramatic 
change in $N_{H}^{(1)}$ at the moment of the flare.  In contrast, 
$N_{H}^{(2)}$, which stays essentially constant throughout the observation, 
decreases drastically exactly at the rise of the flare. After that, it 
remains constant again at $\sim 2.5\times 10^{22}$. Since the 
fluxes shown before were unabsorbed and $N_{H}^{(3)}$ is found to be 
constant during the flare (during the whole observation, in fact), the 
sudden flux increase of component 3 must be intrinsic. On the other hand, 
the flux increase in component~2 could be either intrinsic or due to the 
sudden decrease in $N_{H}^{(2)}$. 

In panels $d$ and $e$ of Fig.~\ref{parm_evol} we show the change in the 
intensities and equivalent width, respectively, of the fluorescence Fe K$\alpha$ 
and Ni K$\alpha$ lines. As can be seen, the intensity of the Fe K$\alpha$ line 
follows the overall brightness of the source well, including the  sharp rise at the 
flares (note again the logarithmic scale in the ordinates). This is clearly seen 
in panel $D$ of Fig.~\ref{corr}. However, the most striking fact is that the 
Fe K$\alpha$ line is clearly less intense after the flare than before. This is also 
clearly seen in panel $D$ of Fig.~\ref{corr} as a hysteresis cycle. The equivalent 
width, in turn, which is related with the density column of the reprocessing material, 
follows the increase in continuum flux up to the flare, when it reaches a value 
of $\sim 230$ eV, and then decreases in the post flare, staying more or less 
constant, at a a value of 180 eV approximately, until the end of the observation 
(panel $E$, Fig.~\ref{corr}). This behaviour must arise from a depletion of neutral 
Fe in the circumsource matter after the flare. We recall that the energy of the 
Fe K$\alpha$ lines remains constant throughout the observation and shows 
no sign of increasing ionization. Therefore, the depletion in neutral Fe 
must reflect a real depletion of circumsource matter. Regarding the Ni K$\alpha$ 
line, its parameters show the same trend as those of the Fe K$\alpha$ line, albeit
with larger uncertainties.


\subsection{Correlations among parameters} 

As mentioned in the previous section, we observe different correlations
among the continuum parameters, looking in detail at panels~B and
C of Fig.~\ref{corr}, three branches are clearly identified:

\begin{itemize}

 \item Before the first peak, all fluxes decrease but component 1 
and 3 do so more rapidly than component 2. 

\item From the first peak to the pre-flare (red to blue points) all
fluxes correlate. 

\item During the flare and later, the fluxes correlate again, but 
with a different slope than before.

\end{itemize}

The same three branches are observed when we plot the Fe K$\alpha$ 
fluorescence line area against the normalization factor of the 
second component (see panel $D$ of Fig.~\ref{corr}). Owing to the
larger uncertainties, there is no clearly visible correlation
when plotting the equivalent width of this line instead 
(panel $E$ of Fig.~\ref{corr}). Still, two different regions are 
observed. During the rise and first peak of the flare the equivalent 
width reaches its highest value, with no clear variation despite 
significant changes in the flux of the second component.
Before and after the flare, there is an apparent correlation
between these two parameters. There is an indication for a different 
correlation coefficient before and after the flare, but the relatively
large uncertainties of the equivalent width do not allow drawing a
firm conclusion. 

\section{Discussion}\label{disc}

\subsection{Source brightness and variability}

To put the current observation into perspective,
it is instructive to compare the obtained fluxes and derived luminosities with 
other observations. Such comparisons, however, depend strongly on 
assumptions of the overall spectral shape, which for many historical data sets
is not well constrained, leading quickly to systematic uncertainties of 
a few 10\% or more. 

For this reason, we assumed the spectrum of Vela X-1 above 10\,keV to be modified by a 
Fermi-Dirac cutoff \citep{tanaka86}, as found by \citet{krey08}, among others. Since the 
luminosity strongly depends on the parameters used to describe that high energy cutoff, we 
decided to take those parameters listed in Table~5 of \citet{krey08}, which 
result in a very soft broad-band spectrum. Choosing another set of parameters 
led to a harder spectrum and an increase in the luminosity estimation
at those energies, which might overestimate the real source luminosity. The 
assumed parameters are therefore 
$E_\mathrm{cut} = 35.6^{+7.5}_{-11.5}$\,keV and $E_\mathrm{fold} = 11.2^{+0.5}_{-0.3}\,keV$. 
The power law photon index $\Gamma$ is consistent with the value presented here 
(see Section~\ref{spe}).

   \begin{figure}
   \centering
   \includegraphics[width=0.95\linewidth]{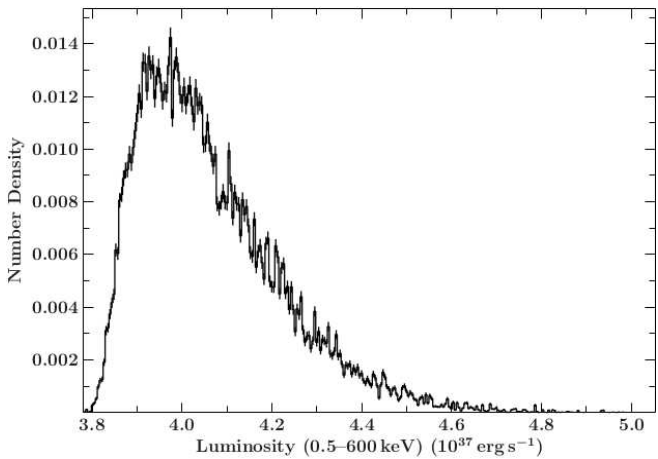}
   \caption{Distribution of X-ray luminosities for the peak of the flare in the energy range 
   \mbox{0.5--600\,keV} using simulated spectra.}
      \label{Lx_estimation}
   \end{figure}

To derive proper values for the uncertainty of the estimated luminosity, we have to 
take the uncertainties of the fit parameters into account, as well as those of the assumed 
FD cutoff. Based on a Monte Carlo approach, we calculated the luminosity of spectrum 56 (the 
peak of the flare) with varying random parameters, corresponding to their individual asymmetric 
uncertainties. The resulting simulated distribution of luminosities between 0.5 and 600\,keV 
after 100000 runs is shown in Figure~\ref{Lx_estimation}. The uncertainty interval of the most-likely 
value was chosen such, that it contains 90\% of the distribution. The final value of 
the peak luminosity between 0.5 and 600 keV of Vela X-1 during the XMM-Newton observation is \newline
$L=3.92^{+0.42}_{-0.09} \times 10^{37}$~erg~s$^{-1}$.

Based on hard X-ray data from INTEGRAL observations, \citet{fuerst2010} found a 
log-normal distribution of flux values, with a median absolute luminosity 
$\langle L_X\rangle = 5.1\times 10^{36}$~erg~s$^{-1}$ and multiplicative
standard deviation $\sigma \approx 2$. Even though a direct comparison 
with \citet{fuerst2010} is difficult because of possible variations in the continuum,
it is clear that the flare around orbital phase 0.22 is brighter 
than $\sim$ 99.9\% of the 3.6~Ms INTEGRAL monitoring data, and thus 
certainly is one of the rare giant flares. 

\subsection{Probable origin of the spectral components}

Although we have used a phenomenological model, the behaviour of the three power law components
in Fig.~\ref{parm_evol} with respect to the flare immediately suggests a further unifying idea of 
only two physical sources: the neutron-star surface and the scattered emission by electrons in the wind.
Continuum components 2 and 3 suffer parameter discontinuities across the flare with significant increases
in the flux relative to component 1 coinciding with a steep decrease in component 2's
absorption such that the post-flare fluxes of
components 1 and 2 are essentially identical during the subsequent decline of about
an order of magnitude in brightness. This suggests that components 1 and 2 are two aspects of the same component that originates at the
neutron-star surface.  Absorption along the line-of-sight through the wind is complex
owing to the warm ionization state of the supergiant's expanding envelope but
can be modelled successfully by the sum of two conventional
cold absorption laws. The balance between these two changed across the flare because photoionization altered the ionic balance
in the wind by removing electrons from the ions responsible for absorption near 1 keV.
These extra free electrons were then responsible for the extra scattering that produced the increase in component 3.

As we have seen, $N_{\rm H}^{(1)}$ decreases from $\geqslant 10^{24}$ at orbital
phase $\phi=0.14$ to $\geqslant 10^{23}$ at orbital phase $\phi=0.26$ as the neutron
star emerges from eclipse and moves along the orbit (Fig.~\ref{parm_evol},
third panel). However, there is clearly a wind density enhancement starting at
$\phi\approx 0.185$, followed by an increase in the emission of the neutron star
due probably to the corresponding gradual accretion of more mass. The later, more sudden increase in
emission from the neutron star was probably due to the accretion of a dense \emph{clump} in the wind, although
this was accompanied by no change of spectral index. The time scale for the rise of the
flare was of the order of one time bin or $t_{\rm rise}\sim 10^{3}$ s.

\subsection{Physical parameters of wind clumps}

Our observation allows us to make direct estimates of the physical properties of the 
possible clump responsible for the giant flare. Considering a spherical
shape of the clump and taking  the characteristic time for the accretion 
of the clump into account would be $t_{\rm rise}\sim 10^{3}$ s. Assuming that the clumps are 
moving with the same velocity as the bulk of the wind, they are travelling at 
$v(r)=v_{\infty}(1-R_{*}/r)^{\beta}=0.46v_{\infty}=5.1\times 10^{7}$ cm/s, 
with $v_{\infty}=1100$ km/s, $r=1.6\,R_{*}$ and $\beta=0.8$. Therefore,
the size of the clump is of the order of

\begin{equation}
l_{\rm cl}\sim t_{\rm rise}v\simeq 5\times 10^{10} \ \rm cm
\end{equation}
or $\sim 0.02R_{*}$, in agreement with theoretical expectations from massive 
star winds (\citet{oskinova11}, Table 1). 

Using a finer time bin, as that used in Fig. 3 (141 s, half the NS spin period) we still see four or five 
data points during the rise of the big flare compatible with a $t_{\rm rise}\lesssim 10^{3}$ s. We can 
now estimate the density of the clump from
\begin{equation}
n_{\rm cl}\sim \frac{N_{\rm H}^{\rm wind}}{l_{\rm cl}}\approx \frac{5\times 10^{23}}{5\times 10^{10}}\approx 10^{13} \ \rm H\  atoms/cm^{3} = 10^{-11} \ \rm g/cm^{3}.
\end{equation}
On the other hand, the characteristic volume of the clump would be 
$V_{cl}\sim {l_{\rm cl}}^{3}\approx 10^{32} \ \rm cm^{3}$. 
Therefore, the mass of the clump responsible of the 
bright flare would be $m_{\rm cl}\sim 10^{21}$~g.

This value is two orders of magnitude greater than deduced by \citeauthor{odaka13} (2013; Eq 13). 
As we have argued, however, the flare we are analysing here is much brighter than 
the one seen by {\it Suzaku}. Our estimate agrees with the one  by \citet{fuerst2010} 
for clumps responsible for bright flares like the one we have observed. We can also estimate  
the luminosity produced by the accretion of such a clump onto the neutron star (NS). Since this 
clump accretes in a characteristic time of $t_{\rm rise}\sim 10^{3}$ s, 
the mass accretion rate will be $\dot{M}\sim 10^{18}$ g/s. Now, if the mass accretion-to-energy 
conversion efficiency $\eta$ adopts reasonable values $[0.1-0.3]$, the luminosity of the flare 
should then be of the order of 

\begin{equation}
L_{X}=\eta G \frac{M_{\rm NS}\dot{M}}{R_{\rm NS}}\approx [2-6]\times 10^{37} \ \rm erg/s
\end{equation} as observed. In conclusion the bright flare observed by {\it XMM-Newton} is consistent 
with the accretion of a dense clump whose size is otherwise the one expected for the structured 
stellar wind of a B0.5Ib star at $r\lesssim 2R_{*}$.

\section{Summary and conclusions}\label{sum}

The observation during eclipse egress has revealed unexpectedly rich behaviour of the Vela X-1 system. 
From the light curve analysis, we found strong variations in source flux (overall and relative flux in
different bands) on all time scales studied from kilo-seconds down to fractions of the pulse period. 
At orbital phase 0.22, a giant flare took place, which was mainly driven by a change in the absorbing and reprocessing 
material between the source and the observer. Even though the observed short-term variability in the light 
curves appears to be mostly dominated by intrinsic flux variations, the spectral shape of the system remains 
rather stable along the observation, reflecting the marked effect of the strong absorption in the 
circumstellar wind on the overall observed count rates. 

The observation was divided into 88 spectra of 1.1 ks exposure time each to study the spectral evolution of the system 
along the orbit. A phenomenological model allows us to fit all spectra well throughout the whole observation,
{\it including the flare}. It consists of three continuum components formed by an absorbed power-law, with the same 
power-law index for each component. This continuum spectral model is complemented by three fluorescent emission lines
of Fe K$_\alpha$, Fe K$_\beta$, and Ni K$_\alpha$. The main results from the spectral analysis follow:

\begin{itemize}

\item  The power-law index remains constant along the observation
with a best value of $\Gamma = 1.595\pm 0.010$.

\item The overall unabsorbed fluxes of components 
2 and 3 increase during the observation, and meanwhile 
the overall unabsorbed flux of component 1 slightly 
decreases from the beginning to the end of the observation. 

\item Two flares are clearly observed at $\phi\approx 0.17$ 
and 0.22 (giant flare) in the unabsorbed fluxes of the
three components.

\item During the rise of the giant flare, which lasted 
$~$1.1 ks, the most dramatic changes were in 
components 3 and 2, with two and one order of magnitude 
increases in their unabsorbed fluxes, respectively. 

\item The absorption column of the first component decreases
from values $\geqslant 10^{24}$ cm$^{-2}$ at the beginning 
of eclipse egress up to $\sim 1.8\times 10^{23}$ cm$^{-2}$ 
at quadrature. Meanwhile, the absorption column of the second
component stays constant from the beginning of the observation
until the rise of the flare, when it decreases drastically from 
$10^{23}$ cm$^{-2}$ to $~$ $2.5 \times 10^{22}$ cm$^{-2}$.

On the other hand, the absorption column of the third 
component remains constant throughout of the observation with 
a best value of $(0.75 \pm 0.03)$ $\times$ 10$^{22}$ cm$^{-2}$, 
in excellent agreement with the ISM absorption towards HD~77581, 
the optical counterpart of Vela X-1.

\item The energy of the Fe K$_\alpha$ line was
constant during the observation with a best value of 
6.435$\pm 0.001$ keV, reflecting no sign of increasing 
ionization, even during the giant flare.

\item The Fe K$_\alpha$ line flux follows the
overall brightness of the source well, is clearly less intense
after the flare. Moreover, the equivalent width of the line
follows the increase in continuum flux up to the flare,
reaching a maximum value of $\sim$ 230 eV, and decreasing
after the flare achieving a constant value of
180 eV. This observed behaviour of the line parameters
clearly indicates a depletion of the neutral Fe
in the circumsource matter after the giant
flare. 

\item The unabsorbed fluxes of three components are 
correlated very well during the whole observation except at the peak of the 
flare. We could establish three different regimes of correlation
among these parameters, which are also observed between
the Fe K$_\alpha$ flux line and the flux of the second component.
 
\end{itemize}

While our results improve understanding of the stellar wind
structure of Vela~X-1 and similar sources, they do not allow 
inferring whether the accreted dense clump of matter is caused by 
clumping in the stellar wind of the massive star, independent of 
the accretion process, or by density perturbations caused
by the passage of the accretor through the dense wind  as found
in hydrodynamical simulations such as those of \citet{Blondin91} and 
\citet{Mano2012}. Further insights could be gained from detailed 
simulations that take all these effects into account in a self-consistent 
manner, but they do not exist yet.

\begin{acknowledgements}
      
This work was supported by the Spanish Ministerio de Ciencia e 	
Innovaci\'on through the projects AYA2010-15431 and AIB2010DE-00054.
It was partly supported by the Bundesministerium f\"ur Wirtschaft und 
Technologie under Deutsches Zentrum f\"ur Luft- und Raumfahrt grant 50OR1113. 
This research was made possible in part by a travel grant from the 
Deutscher Akademischer Austauschdienst. JJRR acknowledges the support by the 
Vicerectorat d'Investigaci\'o, Desenvolupament 
i Innovaci\'o de la Universitat d'Alacant under grant GRE12-35.
The authors acknowledge the help of the International Space Science Institute
at Bern, Switzerland and support by the Faculty of the European Space 
Astronomy Centre. The \texttt{SLXfig} package, developed by John E. Davis, was 
used to produce some of the figures within in this paper.

We thank the anonymous referee whose comments allowed us to improve this paper.

\end{acknowledgements}

\bibliographystyle{aa}
\bibliography{bibliography}

\end{document}